\documentclass
[superscriptaddress,secnumarabic,amssymb,amsmath,nobibnotes,aps,prd,showkeys,showpacs,nofootinbib,onecolumn]{revtex4}%
\usepackage{graphicx}
\usepackage{epstopdf}
\usepackage{epsf}
\usepackage{bm}
\usepackage{amsmath}
\usepackage{amsfonts}
\usepackage{amssymb}%
\providecommand{\U}[1]{\protect\rule{.1in}{.1in}}
\begin{document}
\title{de Sitter and Scaling solutions in a higher-order modified teleparallel theory}
\author{Andronikos Paliathanasis}
\email{anpaliat@phys.uoa.gr}
\affiliation{Instituto de Ciencias F\'{\i}sicas y Matem\'{a}ticas, Universidad Austral de
Chile, Valdivia, Chile}
\affiliation{Institute of Systems Science, Durban University of Technology, Durban 4000,
South Africa}

\begin{abstract}
The existence and the stability conditions for some exact relativistic
solutions of special interest are studied in a higher-order modified
teleparallel gravitational theory. The theory with the use of a Lagrange
multiplier is equivalent with that of General Relativity with a minimally
coupled noncanonical field. The conditions for the existence of de Sitter
solutions and ideal gas solutions in the case of vacuum are studied as also
the stability criteria. Furthermore, in the presence of matter the behaviour
of scaling solutions is given. Finally, we discuss the degrees of freedom of
the field equations and we reduce the field equations in an algebraic
equation, where in order to demonstrate our result we show how this
noncanonical scalar field can reproduce the Hubble function of $\Lambda$-cosmology.

\end{abstract}
\keywords{Cosmology; Modified theories of gravity; Exact Solutions; Teleparallel;}
\pacs{98.80.-k, 95.35.+d, 95.36.+x}
\date{\today}
\maketitle

\section{Introduction}

Modified theories of gravity provide a geometrical mechanism in order to
explain the recent cosmological data \cite{Teg,Kowal,Komatsu,Ade15}.
Precisely, the new terms which are introduced in the gravitational Action
Integral provide components in the field equations which drive the dynamics
and recover the observations. For an extended review on the application of
modified theories of gravity in cosmological studies see \cite{clifton}.

The modification of the gravitational action is not the only possible way to
explain the recent observations. Another proposed approach is to consider, in
the context of General Relativity, the existence of an exotic matter source
such that scalar fields, fluids with particle creation mechanics and others,
for instance see
\cite{fere,Overduin,Barrow,sahni1,Kame,Linder,Lima,sahni,Gal1,barvis,LB10,JOBL11,LBC12,Supri,Wang,suarez,sus}%
.\ Indeed, the two different theoretical approaches to explain the observation
have different origins. However, some modified theories of gravity can be
related with some dark energy models, for instance new degrees of freedom can
be attribute to scalar fields.

From the plethora of proposed modified theories of gravity (see
\cite{mod0,mod00,mod1,mod2,mod3,mod4,mod5,mod6,mod7,mod8} and references
therein) those which have drawn attention in recent years are the $f$-theories
with various applications in all the gravitational areas of study
\cite{Sotiriou,harko,libi,nozi,olmo,Ferraro,Ferraro2,ftSot,ftTam,ftAp,said1,said2,aref7,aref9,aref10,aref11,aref12,aref14,aref15,aref16}%
. In the $f-$theories a generic function of some geometric invariants is
introduced into the Einstein-Hilbert Action Integral. This new term in the
gravitational Action Integral can have phenomenological origin, that is to be
a toy model, or it can have a physical origin such as in the Starobinsky model
for inflation in which quantum-gravitational effects are considered
\cite{star}.

In this work we are interested in the modified $f-$theory in which the
invariants which are used to modify the Einstein-Hilbert Action are the
Ricciscalar~$R$ and the invariant $T$ of the Weitzenb\"{o}ck connection
\cite{myr11,bahamonde}. That theory can be seen as a modification of the
$f\left(  R\right)  ,$ or of the $f\left(  T\right)  $ gravitational theories
in the sense that in our considerations the mentioned theories can be recovered.

Recently, the cosmological evolution for a particular form of that modified
theory was carried out in \cite{anprd}. It was considered that the
gravitational action integral is linear on $T$, or equivalently on $R$, where
the function, $f$, depends upon the boundary term which relates the two
invariants, $R$ and $T$. For that consideration it was found that with the use
of a Lagrange Multiplier a noncanonical scalar field can be introduced in
order to attribute the higher-order derivatives while in the minisuperspace
approach the degrees of freedom are those of general relativity with scalar
field. As far as concerns the dynamical analysis, it was found that this
particular theory can provide two accelerated eras, one stable and one
unstable. \ The unstable accelerated era can be related to the early
acceleration phase (inflation) while the stable accelerated era can correspond
to the late acceleration phase of the Universe. Furthermore it was found that
tracker solutions can exist, that is, the scalar field mimics the matter source.

In this work we consider that the line element which describes the underlying
spacetime is that of the spatially flat FLRW metric and we study the existence
and the stability of two relativistic solutions of special interest. In
particular, we find the condition in which the noncanonical scalar field,
consequently the modified theory, should satisfy in order for de Sitter
solutions to exist, while the stability conditions are determined.
Furthermore, we perform the same analysis for scaling solution with or without
an additional matter source.

Moreover, the theory that we consider does not modify the gravitational
constant and the noncanonical scalar field equivalence is minimally coupled.
Hence, when the scalar field disappears, the theory reduces to standard
General Relativity so that we can say that the theory is defined in the
Einstein frame. Furthermore, as was shown in \cite{anprd}, there exists a
transformation which relates the minisuperspace Lagrangian of that
noncanonical scalar field with the Lagrangian density of a canonical scalar
field. However, that transformation is not a conformal transformation. The
plan of the paper follows.

Section \ref{field} includes the mathematical background for the theory that
we consider. We start by presenting the basic definitions for the teleparallel
equivalence of General Gelativity and we write the field equations for the
theory in which we are interested. Furthermore we define the noncanonical
scalar field which we assume describes the dark energy of the
Universe.\ Finally, we write the modified equations in terms of the classical
form of Friedmann's Equations. In the case of a vacuum in Section
\ref{deSitter} we study the existence and the stability of the de Sitter and
the scaling solutions. Furthermore the stability of scaling solutions in the
presence of matter is studied in Section \ref{scaling}. A discussion on the
conservation laws of constrained Hamiltonian systems is given in Section
\ref{algebraic}. We use that analysis in order to study the integrability of
the field equations for our model and, finally, to reduce the differential
equations in a nonlinear algebraic equation. In Section \ref{conc} we discuss
our results and draw our conclusions.

\section{The modified field equations}

\label{field}

In the theory in which we are interested in the Action integral there exists a
function of second-derivatives for the metric coefficients. It follows that
the gravitational field equations are of fourth-order as in the case of
$f\left(  R\right)  $-gravity. In the following, and for simplicity we follow
the notation of \cite{bahamonde} in which the field equations were derived in
terms of the vierbein field, $e_{i}$, and the boundary term which relates the
two invariants, $R$ and $T$, have been applied.

Let ${\mathbf{e}_{i}(x^{\mu})~}$be the vierbein fields, as nonholonomic frames
in spacetime, which are the dynamical variables of teleparallel gravity. The
vierbein fields form an orthonormal basis for the tangent space at each point,
$x^{\mu}$, of the manifold, that is, $g(e_{i},e_{j})=\mathbf{e}_{i}%
\cdot\mathbf{e}_{i}=\eta_{ij}$, where $\eta_{ij}~$is the line element of
four-dimensional Minkowski spacetime. In a coordinate basis~the vierbeins can
be written as $e_{i}=h_{i}^{\mu}\left(  x\right)  \partial_{i},$ for which the
the metric tensor is defined as follows%
\begin{equation}
g_{\mu\nu}(x)=\eta_{ij}h_{\mu}^{i}(x)h_{\nu}^{j}(x). \label{ft.01}%
\end{equation}

The curvatureless Weitzenb\"{o}ck connection $\hat{\Gamma}^{\lambda}{}_{\mu
\nu}=h_{a}^{\lambda}\partial_{\mu}h_{\nu}^{a}$ has the nonnull torsion tensor,
\cite{ftt0,ftt1}
\begin{equation}
T_{\mu\nu}^{\beta}=\hat{\Gamma}_{\nu\mu}^{\beta}-\hat{\Gamma}_{\mu\nu}^{\beta
}=h_{i}^{\beta}(\partial_{\mu}h_{\nu}^{i}-\partial_{\nu}h_{\mu}^{i}),
\label{ft.02}%
\end{equation}
while the Lagrangian density of the teleparallel gravity, from which the
gravitational field equations are derived, is the scalar $\ T={S_{\beta}}%
^{\mu\nu}{T^{\beta}}_{\mu\nu},~$where ${S_{\beta}}^{\mu\nu}~$is defined as
\begin{equation}
{S_{\beta}}^{\mu\nu}=\frac{1}{2}({K^{\mu\nu}}_{\beta}+\delta_{\beta}^{\mu
}{T^{\theta\nu}}_{\theta}-\delta_{\beta}^{\nu}{T^{\theta\mu}}_{\theta}).
\label{ft.03}%
\end{equation}
Furthermore the geometric quantity ${K^{\mu\nu}}_{\beta}$ is called the
contorsion tensor and equals the difference between the Levi-Civita
connections in the holonomic and the nonholonomic frame and it is defined by
the nonnull torsion tensor, ${T^{\mu\nu}}_{\beta}$, as
\begin{equation}
{K^{\mu\nu}}_{\beta}=-\frac{1}{2}({T^{\mu\nu}}_{\beta}-{T^{\nu\mu}}_{\beta
}-{T_{\beta}}^{\mu\nu}). \label{ft.03b}%
\end{equation}

Consider now the gravitational Action Integral to be%

\begin{equation}
S\equiv\frac{1}{16\pi G}\int d^{4}xe\left[  f(T,R+T)\right]  +S_{m}\equiv
\frac{1}{16\pi G}\int d^{4}xe\left[  f(T,B)\right]  +S_{m}, \label{ftb.01}%
\end{equation}
in which $e=\det(e_{\mu}^{i})=\sqrt{-g},~S_{m}$ is the Action Integral for the
matter source and $B$ is the boundary term $B=2e_{\nu}^{-1}\partial_{\nu
}\left(  eT_{\rho}^{~\rho\nu}\right)  .$

The gravitational field equations are derived to be%
\begin{align}
4\pi Ge\mathcal{T}_{a}^{\left(  m\right)  }{}^{\lambda}  &  =\frac{1}{2}%
eh_{a}^{\lambda}\left(  f_{,B}\right)  ^{;\mu\nu}g_{\mu\nu}-\frac{1}{2}%
eh_{a}^{\sigma}\left(  f_{,B}\right)  _{;\sigma}^{~~~;\lambda}+\frac{1}%
{4}e\left(  Bf_{,B}-\frac{1}{4}f\right)  h_{a}^{\lambda}\,+(eS_{a}{}%
^{\mu\lambda})_{,\mu}f_{,T}\nonumber\\
&  ~\ ~+e\left(  (f_{,B})_{,\mu}+(f_{,T})_{,\mu}\right)  S_{a}{}^{\mu\lambda
}~-ef_{,T}T^{\sigma}{}_{\mu a}S_{\sigma}{}^{\lambda\mu}, \label{ftb.02}%
\end{align}
where $\mathcal{T}_{a}^{\left(  m\right)  }{}^{\lambda}$ is the
energy-momentum tensor of the matter source and as usual comma denotes partial
derivative, while \textquotedblleft$;$\textquotedblright\ $\ $ denotes
covariant derivative. At the same time it is straightforward to observe that,
when $f_{,BB}=0$, the field equations reduce to those of $f\left(  T\right)  $
teleparallel gravity and for $f_{,BB}\neq0$ the theory is of fourth-order as
in the boundary term second-order derivatives exist.

We can rewrite the field equations (\ref{ftb.02}) by using the Einstein
tensor, $G_{a}^{\lambda},~$as
\begin{align}
4\pi Ge\mathcal{T}_{a}^{\left(  m\right)  }{}^{\lambda}  &  =ef_{,T}%
G_{a}^{\lambda}+\left[  \frac{1}{4}\left(  Tf_{,T}-f\right)  eh_{a}^{\lambda
}+e(f_{,T})_{,\mu}S_{a}{}^{\mu\lambda}\right]  +\label{ftb.04}\\
&  +\left[  e(f_{,B})_{,\mu}S_{a}{}^{\mu\lambda}-\frac{1}{2}e\left(
h_{a}^{\sigma}\left(  f_{,B}\right)  _{;\sigma}^{~~~;\lambda}-h_{a}^{\lambda
}\left(  f_{,B}\right)  ^{;\mu\nu}g_{\mu\nu}\right)  +\frac{1}{4}%
eBh_{a}^{\lambda}f_{,B}\right] \nonumber
\end{align}
or equivalently in the following form%
\begin{equation}
ef_{,T}G_{a}^{\lambda}=4\pi Ge\mathcal{T}_{a}^{\left(  m\right)  }{}^{\lambda
}+4\pi Ge\mathcal{T}_{a}^{\left(  DE\right)  }{}^{\lambda}, \label{ftb.05}%
\end{equation}
where~$\mathcal{T}_{a}^{\left(  DE\right)  }{}^{\lambda}$ is the effective
energy momentum tensor which attributes the additional dynamical terms which
follows from the modified Action Integral, that is,
\begin{align}
4\pi Ge\mathcal{T}_{a}^{\left(  DE\right)  }{}^{\lambda}  &  =-\left[
\frac{1}{4}\left(  Tf_{,T}-f\right)  eh_{a}^{\lambda}+e(f_{,T})_{,\mu}S_{a}%
{}^{\mu\lambda}\right]  +\label{ftb.06}\\
&  -\left[  e(f_{,B})_{,\mu}S_{a}{}^{\mu\lambda}-\frac{1}{2}e\left(
h_{a}^{\sigma}\left(  f_{,B}\right)  _{;\sigma}^{~~~;\lambda}-h_{a}^{\lambda
}\left(  f_{,B}\right)  ^{;\mu\nu}g_{\mu\nu}\right)  +\frac{1}{4}%
eBh_{a}^{\lambda}f_{,B}\right]  .\nonumber
\end{align}

Equation (\ref{ftb.05}) can be written as
\begin{equation}
eG_{a}^{\lambda}=G_{eff}\left(  e\mathcal{T}_{a}^{\left(  m\right)  }%
{}^{\lambda}+e\mathcal{T}_{a}^{\left(  DE\right)  }{}^{\lambda}\right)  ,
\label{ftb.07}%
\end{equation}
in which
\begin{equation}
G_{eff}=\frac{4\pi G}{f_{,T}}, \label{ftb.08}%
\end{equation}
is an effective varying gravitational constant.

As far the effective energy momentum tensor, $\mathcal{T}_{a}^{\left(
DE\right)  }{}^{\lambda}$,V is concerned, it can be seen that one can define
two components, one which has its origins on the teleparallel part while the
second part on the boundary term. Hence we write%
\begin{equation}
\mathcal{T}_{a}^{\left(  DE\right)  }{}^{\lambda}=\mathcal{T}_{a}^{\left(
B\right)  }{}^{\lambda}+\mathcal{T}_{a}^{\left(  B\right)  }{}^{\lambda},
\label{ftb.09}%
\end{equation}
where now%
\begin{equation}
4\pi Ge\mathcal{T}_{a}^{\left(  T\right)  }{}^{\lambda}=-\left[  \frac{1}%
{4}\left(  Tf_{,T}-f\right)  eh_{a}^{\lambda}+e(f_{,T})_{,\mu}S_{a}{}%
^{\mu\lambda}\right]  \label{ftb.10}%
\end{equation}
and%
\begin{equation}
4\pi Ge\mathcal{T}_{a}^{\left(  B\right)  }{}^{\lambda}=-\left[
e(f_{,B})_{,\mu}S_{a}{}^{\mu\lambda}-\frac{1}{2}e\left(  h_{a}^{\sigma}\left(
f_{,B}\right)  _{;\sigma}^{~~~;\lambda}-h_{a}^{\lambda}\left(  f_{,B}\right)
^{;\mu\nu}g_{\mu\nu}\right)  +\frac{1}{4}eBf_{,B}h_{a}^{\lambda}\right]  .
\label{ftb.11}%
\end{equation}

There is a special case in which the function $f\left(  T,B\right)  $ can be
written as the sum of two functions, such that $f\left(  T,B\right)  =F\left(
T\right)  +\Phi\left(  B\right)  $. Then expressions, (\ref{ftb.10}) and
(\ref{ftb.11}), are simplified as%
\begin{equation}
4\pi Ge\mathcal{T}_{a}^{\left(  T\right)  }{}^{\lambda}=-\left[  \frac{1}%
{4}\left(  TF_{,T}-F\right)  eh_{a}^{\lambda}+e(F_{,TT})T_{,\mu}S_{a}{}%
^{\mu\lambda}\right]  \label{ftb.12}%
\end{equation}
and%
\begin{equation}
4\pi Ge\mathcal{T}_{a}^{\left(  B\right)  }{}^{\lambda}=-\left[  e(\Phi
_{,BB})B_{;\mu}S_{a}{}^{\mu\lambda}-\frac{1}{2}e\left(  h_{a}^{\sigma}\left(
\Phi_{,B}\right)  _{;\sigma}^{~~~;\lambda}-h_{a}^{\lambda}\left(  \Phi
_{,B}\right)  ^{;\mu\nu}g_{\mu\nu}\right)  +\frac{1}{4}e\left(  B\Phi
_{,B}-\Phi\right)  h_{a}^{\lambda}\right]  \label{ftb.13}%
\end{equation}
while the varying gravitational constant is $\frac{G_{eff}}{4\pi G}=\left(
F_{,T}\right)  ^{-1}.~$\ 

\section{The $f\left(  T,B\right)  =T+\Phi\left(  B\right)  $ theory}

The theory that we consider is the one in which $F\left(  T\right)  $ is
linear and $\Phi_{,BB}\neq0.$ For that special consideration the field
equations take the simple form\footnote{Without loss of generality we assume
that $F\left(  T\right)  =T.$}%
\begin{equation}
G_{a}^{\lambda}=4\pi G\left(  e\mathcal{T}_{a}^{\left(  m\right)  }{}%
^{\lambda}+e\mathcal{T}_{a}^{\left(  B\right)  }{}^{\lambda}\right)  ,
\label{ftb.14}%
\end{equation}
where only the $\mathcal{T}_{a}^{\left(  B\right)  }{}^{\lambda}$ tensor
survives and contributes to the dark sector of the universe while the
gravitational constant remains constant, something which is not true in the
pure $f\left(  T\right)  $ or $f\left(  R\right)  $ theories of gravity and an
effective gravitational constant is defined.

That specific case, $f\left(  T,B\right)  =T+\Phi\left(  B\right)  ,$ was the
main subject of study in \cite{anprd} in the cosmological scenario of a
spatially flat FLRW spacetime. Moreover, with the use of Lagrange Multipliers
the extra degrees of freedom have been attributed to annoncanonical scalar
field. \ In the following we assume dimensions such that $4\pi G=1$.

Consider the spatially flat FLRW universe with line element%
\begin{equation}
ds^{2}=-N^{2}\left(  t\right)  dt^{2}+a^{2}\left(  t\right)  \left(
dx^{2}+dy^{2}+dz^{2}\right)  , \label{ftb.15}%
\end{equation}
where $a\left(  t\right)  $ is the scale factor and describes the radius of
the three-dimensional Euclidean space and $N\left(  t\right)  $ is the lapse
function. Furthermore from the cosmological principle we select the observer
to be $u^{\mu}=\frac{1}{N}\delta_{t}^{\mu}~$such that $u^{\mu}u_{\mu}=-1$.

Furthermore for the vierbein we consider the following diagonal frame%
\begin{equation}
h_{\mu}^{i}(t)=\left(
\begin{tabular}
[c]{cccc}%
$N\left(  t\right)  $ &  &  & \\
& $a\left(  t\right)  $ &  & \\
&  & $a\left(  t\right)  $ & \\
&  &  & $a\left(  t\right)  $%
\end{tabular}
\ \ \ \ \ \right)  , \label{ftb.150}%
\end{equation}
from which it follows that%
\begin{equation}
T=-\frac{6}{N^{2}}\left(  \frac{\dot{a}}{a}\right)  ^{2}~,~B=-\frac{6}{N^{2}%
}\left(  \frac{\ddot{a}}{a}+\frac{2\dot{a}^{2}}{a^{2}}-\frac{\dot{a}\dot{N}%
}{aN}\right)  .\, \label{ftb.15a}%
\end{equation}

Therefore, with the use of the Lagrange Multiplier, the gravitational field
equations (\ref{ftb.07}) follow from the variation of the\ Action with
Lagrangian
\begin{equation}
\mathcal{L}\left(  N,a,\dot{a},B,\dot{B}\right)  =-\frac{6}{N}a\dot{a}%
^{2}+\frac{6}{N}a^{2}\Phi\left(  B\right)  _{,BB}\dot{a}\dot{B}+Na^{3}\left(
\Phi\left(  B\right)  -B\Phi\left(  B\right)  _{,B}\right)  +matter
\label{ftb.15b}%
\end{equation}
or equivalently
\begin{equation}
\mathcal{L}\left(  N,a,\dot{a},\phi,\dot{\phi}\right)  =-\frac{6}{N}a\dot
{a}^{2}+\frac{6}{N}a^{2}\dot{a}\dot{\phi}-Na^{3}V\left(  \phi\right)  +matter,
\label{ftb.16}%
\end{equation}
where the higher-order derivatives have been attributed to the noncanonical
field
\begin{equation}
\phi=\Phi\left(  B\right)  _{,B}~,~V\left(  \phi\right)  =B\Phi\left(
B\right)  _{,B}-\Phi\left(  B\right)  , \label{ftb.17}%
\end{equation}
as an analogy of other higher-order theories of gravities.

Furthermore from (\ref{ftb.13}) by using (\ref{ftb.17}) we have that the
energy momentum tensor of the noncanonical scalar field is
\begin{equation}
4\pi Ge\mathcal{T}_{a}^{\left(  \phi\right)  }{}^{\lambda}=\left[  \frac{1}%
{2}e\left(  h_{a}^{\sigma}\left(  \phi\right)  _{;\sigma}^{~~~;\lambda}%
-h_{a}^{\lambda}\left(  \phi\right)  ^{;\mu\nu}g_{\mu\nu}\right)  -e\phi
_{;\mu}S_{a}{}^{\mu\lambda}-\frac{1}{4}eV\left(  \phi\right)  h_{a}^{\lambda
}\right]  . \label{ftb.17a}%
\end{equation}

\subsection{Dark energy fluid components}

In the case of the lapse function, $N\left(  t\right)  =1$, the field
equations are derived to be
\begin{equation}
3H^{2}=3H\dot{\phi}+\frac{1}{2}V\left(  \phi\right)  +\rho_{m}, \label{ftb.18}%
\end{equation}%
\begin{equation}
2\dot{H}+3H^{2}=\ddot{\phi}+\frac{1}{2}V\left(  \phi\right)  -p_{m}
\label{ftb.19}%
\end{equation}
and the constraint equation
\begin{equation}
\frac{1}{6}V_{,\phi}+\dot{H}+3H^{2}=0 \label{ftb.20}%
\end{equation}
which is nothing else than the definition of the boundary term, $B$. Note that
$B=V_{,\phi}$. Finally for the matter source we assume that is minimally
coupled with the theory which means that the differential equation,
\begin{equation}
\dot{\rho}_{m}+3H\left(  \rho_{m}+p_{m}\right)  =0, \label{ftb.20a}%
\end{equation}
holds.

The field equations, (\ref{ftb.18}) and\ (\ref{ftb.19}), can be written as%
\begin{equation}
3H^{2}=\rho_{DE}^{\left(  B\right)  }+\rho_{m} \label{ftb.21}%
\end{equation}
and
\begin{equation}
2\dot{H}+3H^{2}=-p_{DE}^{\left(  B\right)  }-p_{m}, \label{ftb.22}%
\end{equation}
where $\rho_{DE}^{\left(  B\right)  }$ and $p_{DE}^{\left(  B\right)  }$ ~are
the energy density and the pressure components of $\mathcal{T}_{a}^{\left(
B\right)  }{}^{\lambda}$, respectivelt, defined as%

\begin{equation}
\rho_{DE}^{\left(  B\right)  }=3H\dot{\phi}+\frac{1}{2}V\left(  \phi\right)
~\mbox{\rm and}~p_{DE}^{\left(  B\right)  }=-\left(  \ddot{\phi}+\frac{1}%
{2}V\left(  \phi\right)  \right)  , \label{ftb.23}%
\end{equation}
from which we can see that equation (\ref{ftb.20}) can be written as the
continuity equation,
\begin{equation}
\dot{\rho}_{DE}^{\left(  B\right)  }+3H\left(  \rho_{DE}^{\left(  B\right)
}+p_{DE}^{\left(  B\right)  }\right)  =0. \label{ftb.24}%
\end{equation}

Hence the equation of state parameter for the geometric dark energy fluid is
defined as%
\begin{equation}
w_{DE}=\frac{p_{DE}^{\left(  B\right)  }}{\rho_{DE}^{\left(  B\right)  }%
}=-\frac{\ddot{\phi}+\frac{1}{2}V\left(  \phi\right)  }{3H\dot{\phi}+\frac
{1}{2}V\left(  \phi\right)  } \label{ftb.25}%
\end{equation}
while we observe that, when $V\left(  \phi\right)  $ dominates, that is,
$\ddot{\phi}\ll V\left(  \phi\right)  $ and $\dot{\phi}\ll V\left(
\phi\right)  $, the field $\phi$ describes the cosmological constant, that is,
$w_{DE}=-1.$

\section{Stability of special solutions}

\label{deSitter}

In this Section we study the conditions which the theory and function,
$\Phi\left(  B\right)  ,$ should satisfy in order that two classical
solutions, that of the de Sitter Universe and the ideal gas solution, are
recovered by the modified theory. Moreover the stability conditions of these
solutions are determined.

In the spirit of Barrow and Ottewill \cite{barott} in order to perform the
stability analysis we prefer to write the field equations as a set of
higher-order equations. Hence, in the case of vacuum, i.e., $\rho_{m}=p_{m}%
=0$, the field equations (\ref{ftb.18})-(\ref{ftb.20}) can be written
equivalently as%

\begin{align}
\rho_{m}  &  =6a^{2}\dot{a}\left(  \left(  1+2\Phi_{,B}\right)  \dot{a}%
+6\Phi_{,BB}a^{\left(  3\right)  }\right)  +\nonumber\\
&  +6a\ddot{a}\left(  \Phi_{,B}a^{2}+18\Phi_{,BB}\left(  \dot{a}\right)
^{2}\right)  -144\Phi_{,BB}\left(  \dot{a}\right)  ^{4}+\Phi a^{4},
\label{fttb.26}%
\end{align}
and\qquad%
\begin{align}
0  &  =2a^{5}\left(  \left(  2+3\Phi_{,B}\right)  \ddot{a}+6\Phi
_{,BB}a^{\left(  4\right)  }\right)  +72a^{2}\left(  \dot{a}\right)
^{2}\left(  2\Phi_{,BB}\left(  \dot{a}\right)  ^{2}-\Phi_{,BBB}\left(
9\left(  \ddot{a}\right)  ^{2}-8\dot{a}a^{\left(  3\right)  }\right)  \right)
\nonumber\\
&  +2a^{4}\left(  \left(  1+6\Phi_{,B}\right)  \dot{a}^{2}+18\Phi_{,BB}\left(
\ddot{a}\right)  ^{2}+12a^{\left(  3\right)  }\left(  \Phi_{,BB}\dot{a}%
-3\Phi_{,BBB}a^{\left(  3\right)  }\right)  \right)  +\nonumber\\
&  -216a^{3}\dot{a}\ddot{a}\left(  \Phi_{,BB}\dot{a}+2\Phi_{,BBB}a^{\left(
3\right)  }\right)  +1783\Phi_{,BBB}a\left(  \dot{a}\right)  ^{4}\ddot{a}+\Phi
a^{6}+p_{m}, \label{fttb.27}%
\end{align}
or in terms of the Hubble function as
\begin{equation}
0=36\Phi_{,BB}H\ddot{H}+6H^{2}\left(  1+3\Phi_{,B}+36\Phi_{,BB}\dot{H}\right)
+6\Phi_{,B}\dot{H}+\Phi\label{fttb.28}%
\end{equation}
and%
\begin{align}
0  &  =12\Phi_{,BB}H^{\left(  3\right)  }+72\ddot{H}\left(  \Phi_{,BB}%
H-\Phi_{,BBB}\left(  12H\dot{H}-\ddot{H}\right)  \right)  +\Phi+\nonumber\\
&  6H^{2}\left(  1+3\Phi_{,B}-432\Phi_{,BBB}\left(  \dot{H}\right)
^{2}\right)  +2\dot{H}\left(  2+3\Phi_{,B}+36\Phi_{,BB}\dot{H}\right)  ,
\label{fttb.29}%
\end{align}
where $B=-6\left(  \dot{H}+3H^{2}\right)  $.

As in the case of General Relativity the sets of equations (\ref{fttb.26}),
(\ref{fttb.27}) or (\ref{fttb.28}), (\ref{fttb.29}) are not independent. In
particular equations (\ref{fttb.27}), (\ref{fttb.29}) are the total
derivatives of equations (\ref{fttb.26}), (\ref{fttb.28}), respectively.
\ Because of the latter property, for our analysis we select to work directly
with the second-order differential equation, (\ref{fttb.28}).

\subsection{de Sitter Universe}

Consider now that the scale factor is that of the de Sitter Universe, that is,
$a\left(  t\right)  =a_{0}e^{H_{0}t}$. Substitute the latter solution into
(\ref{fttb.28}). Then
\begin{equation}
3\left(  B_{0}\Phi_{,B_{0}}-\Phi\right)  +B_{0}=0, \label{fttb.30}%
\end{equation}
where $B_{0}=-18\left(  H_{0}\right)  ^{2}.$

Condition (\ref{fttb.30}) is the analogue of the Barrow-Ottewill condition of
$f\left(  R\right)  $-gravity. Hence for any functional form of $\Phi\left(
B\right)  $ in which there exists $B_{0}$ such that (\ref{fttb.30}) be true.
Then at the moment $t_{0}~$in which $B\left(  t_{0}\right)  =B_{0}$ the
spacetime is maximally symmetric.

However, there is a family of theories in which the condition (\ref{fttb.30})
is satisfied identically.\ In particular the theory with $\Phi\left(
B\right)  =-\frac{B}{3}\ln B+\Phi_{0}B$ is satisfied for any $B$ condition
(\ref{fttb.30}) as an analogue of the quadratic $f\left(  R\right)  =R^{2}$
theory. However, it is important to mention that, as any linear term of $B$
can be neglected from the Action Integral, we can select without loss of
generality $\Phi_{0}=0$, that is, the theory which satisfies identically
condition (\ref{fttb.30}) is the
\begin{equation}
\Phi\left(  B\right)  =-\frac{B}{3}\ln\left(  B\right)  . \label{fttb.30a}%
\end{equation}

In the scalar field description we find that the potential, $V\left(
\phi\right)  $, which corresponds to the theory (\ref{fttb.30a}) has the
simple exponential form $V\left(  \phi\right)  \simeq e^{-3\phi}$. \ In
\cite{angrg} the Barrow-Ottewill condition was written in terms of the
Brans-Dicke scalar field which is equivalent to $f\left(  R\right)  $-gravity.
In a similar way condition (\ref{fttb.30}) in terms of the scalar field and
the scalar field potential is expressed as%

\begin{equation}
3V\left(  \phi_{0}\right)  +\frac{dV\left(  \phi\right)  }{d\phi}%
|_{\phi\rightarrow\phi_{0}}=0. \label{fttb.30b}%
\end{equation}

In order to study the stability of the de Sitter Universe we consider a small
perturbation in the metric such that $a\left(  t\right)  =a_{0}e^{H_{0}%
t}+\varepsilon a_{P}\left(  t\right)  $, where $\varepsilon$ is an
infinitesimal parameter so that $\varepsilon^{2}\rightarrow0$. That is
equivalent with the perturbation in the Hubble function%
\begin{equation}
H\left(  t\right)  =H_{0}+\frac{\varepsilon}{a_{0}}e^{-H_{0}t}\left(  \dot
{a}_{P}-H_{0}a_{P}\right)  +O\left(  \varepsilon^{2}\right)  \label{fttb.31}%
\end{equation}
or equivalently in the first-order approximation $H\left(  t\right)
=H_{0}+\varepsilon H_{P}\left(  t\right)  $.

Therefore, keeping first-order corrections of the field equation,
(\ref{fttb.28}), we obtain the linear second-order differential equation%
\begin{equation}
3\Phi_{,B_{0}B_{0}}\left(  \ddot{H}_{P}+3H_{0}\dot{H}_{P}\right)  +\left(
1-54H_{0}^{2}\Phi_{,B_{0}B_{0}}\right)  H_{P}=0, \label{fttb.32}%
\end{equation}
where $B_{0}$ is the solution of the algebraic equation, (\ref{fttb.30}).
Moreover it is of special interest to mention that in the linearized equation
only second derivatives of the function $\Phi\left(  B\right)  $ are involved
in contrast to $f\left(  R\right)  $ gravity in which first derivatives of the
function which defines the theory exists.

The analytic solution of the linear equation, (\ref{fttb.32}), can be written
easily in closed-form. We have three different conditions which we should
study. The conditions are $Con\left[  1\right]  \equiv\Phi_{,B_{0}B_{0}}\neq
0$,$~~Con\left[  2\right]  \equiv\left(  1-54H_{0}^{2}\Phi_{,B_{0}B_{0}%
}\right)  \neq0~$and $Con\left[  3\right]  \equiv~\Phi_{,B_{0}B_{0}}+\frac
{8}{27B_{0}}\neq0.$

\begin{itemize}
\item In the most general case in which $Con\left[  1\right]  \neq
0,~Con\left[  2\right]  \neq0$ and $Con\left[  3\right]  \neq0~$the analytic
solution for the perturbations is%
\begin{equation}
H_{P}\left(  t\right)  =H_{P}^{1}\exp\left(  \mu_{+}~t\right)  +H_{P}^{2}%
\exp\left(  \mu_{-}~t\right)  , \label{fttb.33}%
\end{equation}
in which $H_{p}^{1,2}$ are constants of integration and
\begin{equation}
\mu_{\pm}=-\frac{3}{2}H_{0}\pm\frac{\left\vert H_{0}\right\vert }{2}\left(
\frac{243H_{0}^{2}\Phi_{,B_{0}B_{0}}-4}{\Phi_{,B_{0}B_{0}}}\right)  ^{1/2}.
\label{fttb.34}%
\end{equation}
We assume that $t>0$. Hence, when $\mu_{+}$ and $\mu_{-}$ have both negative
real parts, the perturbations decay and the de Sitter Universe is stable. That
is possible when the de Sitter solution describes an expanding Universe, that
is, $H_{0}>0$ and $\Phi_{,B_{0}B_{0}}$ is constrained as follows%
\begin{equation}
-\frac{8}{27B_{0}}<\Phi_{,B_{0}B_{0}}<-\frac{1}{3B_{0}}~\text{~with
}\operatorname{Im}\left(  \mu_{\pm}\right)  =0, \label{fttb.35}%
\end{equation}
or%
\begin{equation}
\Phi_{,B_{0}B_{0}}<-\frac{8}{27B_{0}}~\text{with }\operatorname{Im}\left(
\mu_{\pm}\right)  \neq0. \label{fttb.36}%
\end{equation}
Furthermore, when $H_{0}<0$, the real part of the $\mu_{+}$ is always
positive, $\operatorname{Re}\left(  \mu_{+}\right)  >0$, which means that the
de Sitter solution is stable for initial conditions such that~$H_{P}^{1}=0$
and $\mu_{-}<0$. The latter condition becomes $\Phi_{,B_{0}B_{0}}>-\frac
{1}{3B_{0}}.$

\item We now assume that $Con\left[  1\right]  \neq0,~Con\left[  2\right]
\neq0$ and the third condition vanishes, $Con\left[  3\right]  =0$. In that
case the solution \ of the perturbations is
\begin{equation}
H_{P}\left(  t\right)  =H_{P}^{0}\exp\left(  -\frac{3}{2}H_{0}~t\right)
\left(  t-t_{0}\right)  , \label{fttb.37}%
\end{equation}
which means that the Hubble function is%
\begin{equation}
H\left(  t\right)  =H_{0}+\varepsilon H_{P}^{0}\exp\left(  -\frac{3}{2}%
H_{0}~t\right)  \left(  t-t_{0}\right)  , \label{fttb.38}%
\end{equation}
where now it is easy to see that for $t>0$, when $H_{0}>0$ the de Sitter
solution is always a stable solution.

\item Furthermore, when$~Con\left[  2\right]  =0$, that is, $Con\left[
1\right]  \neq0,~Con\left[  3\right]  \neq0$, from the differential equation
(\ref{fttb.32}) it follows that%
\begin{equation}
H_{p}\left(  t\right)  =-\frac{H_{P}^{1}}{3H_{0}}e^{-3H_{0}t}+H_{P}^{2}.
\label{fttb.39}%
\end{equation}
Hence again for $H_{0}>0$ the de Sitter Universe is a future stable solution.

\item The last case that we have to consider is when $Con\left[  1\right]
=0$, that is,~$\Phi_{,B_{0}B_{0}}=0$ which means that curvature of function
$\Phi\left(  B\right)  $ vanishes at $B_{0}$. There, in this specific case,
the differential equation (\ref{fttb.32}) reduces to the algebraic equation
$H_{P}\left(  t\right)  =0$. Hence the de Sitter solution is always stable.
\end{itemize}

We demonstrate our results with an illustrate example.\ \ Consider the theory,
$\Phi_{A}\left(  B\right)  =\Phi_{0}H_{0}^{2}B^{2}+\Phi_{1}H_{0}^{2}$, which
from (\ref{fttb.30}) we find there exists a de Sitter solution if and only if
$\Phi_{0}=\frac{6+\Phi_{1}}{324H_{0}^{4}}$ or $B_{0}^{\left(  \pm\right)
}=\pm\sqrt{\frac{6+\Phi_{1}}{\Phi_{0}}}$ which means that there exist two
possible de Sitter solutions. Moreover, because $B_{0}=-18H_{0}^{2}$ when
$B_{0}^{\left(  \pm\right)  }<0$, $H_{0}$~is real; while, when $B_{0}^{\left(
\pm\right)  }>0$,$~H_{0}$ is imaginary and we have a bounced universe.

We assume that $\Phi_{1}\neq-6$ which is equivalent with the $\Phi_{0}\neq0,$
and consequently $Con\left[  1\right]  \neq0$. In that scenario we find that
the de Sitter solutions are stable and the perturbations decay when $H_{0}>0$
and $-6<\Phi_{1}<-3$, whereas for $\Phi_{1}<-\frac{10}{3}$ we observe that
$\operatorname{Im}\left(  \mu_{\pm}\right)  \neq0$, while in the case in which
$\Phi_{1}=-\frac{10}{3}$ the expression for the perturbation is given by
(\ref{fttb.40}).

\begin{figure}[ptb]
\includegraphics[height=12.5cm]{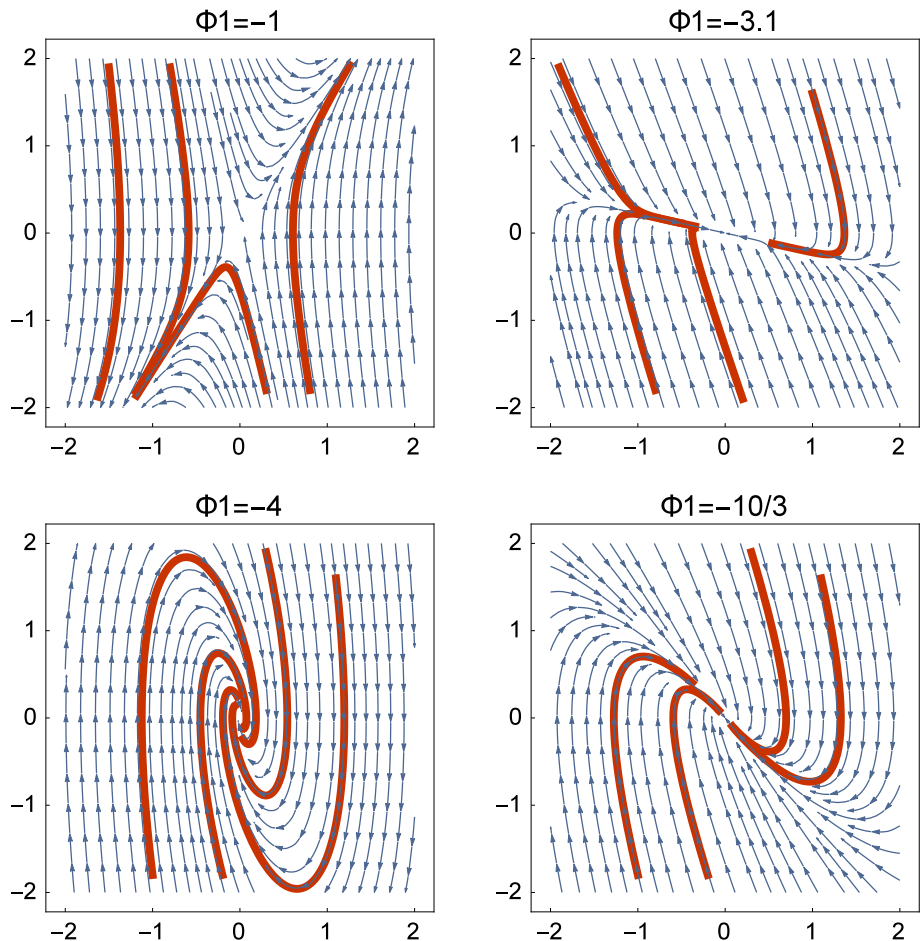}\centering
\caption{Phase portrait for the perturbation equation (\ref{fttb.32}) \ for
$\Phi\left(  B\right)  =\Phi_{0}H_{0}^{2}B^{2}+\Phi_{1}H_{0}^{2}$ $\ $and
$H_{0}>0$. The figs. are for $\Phi_{1}=-1,~\Phi_{1}=-3.1,~\Phi_{1}=-4$ and
$\Phi_{1}=-\frac{10}{3}$.~The red describes solutions of equation
(\ref{fttb.32}). In all figs. the critical point is the $\left(  0,0\right)
$. }%
\label{plot1}%
\end{figure}

In Fig. \ref{plot1} the phase space diagrams are presented for the
perturbation equation (\ref{fttb.32}) for different values of the parameter
$\Phi_{1}$. \ 

Finally we study the stability of the de Sitter solution for the theory
(\ref{fttb.30a}). We observe that the solution for the equation of the
perturbations corresponds to the case with $Con\left[  2\right]  =0$, which
means that the de Sitter solution is always stable when the de Sitter solution
describes an expanding universe, that is, the Hubble constant is positive.


\subsection{Ideal gas solution}

We now assume that the scale factor describes an ideal gas cosmological
solution, that is, $a\left(  t\right)  =a_{0}t^{p}$, where~the equation of
state parameter for the cosmological fluid is $\gamma-1=\frac{2}{3p}$.
Furthermore we assume $\gamma\in(0,2]$, that is, $p\succeq\frac{2}{3}$.

Furthermore in the case of vacuum from the dynamical system, (\ref{ftb.18}%
)-(\ref{ftb.20}), we find that the power-law solution is a closed-form
solution if%

\begin{equation}
\phi\left(  t\right)  =\frac{p}{1+3p}\ln V_{0}+\frac{2p}{1+3p}\ln\left(
\left(  1+3p\right)  t\right)  ~,~V\left(  t\right)  =\frac{\left(
3p-1\right)  }{\left(  3p+1\right)  }\frac{6p^{2}}{t^{2}}, \label{fttb.40}%
\end{equation}
which means that the scalar field potential has the functional form,%

\begin{equation}
V\left(  \phi\right)  =V_{0}6p^{2}\left(  9p^{2}-1\right)  \exp\left(
-\frac{1+3p}{p}\phi\right)  . \label{fttb.41}%
\end{equation}
Therefore from the Clairaut equation (\ref{ftb.17}) it follows that the
corresponding $\Phi\left(  B\right)  $ theory is%
\begin{equation}
\Phi\left(  B\right)  =-\frac{p}{1+3p}B\ln\left(  B\right)  . \label{fttb.42}%
\end{equation}

Potential, (\ref{fttb.41}), is not the only possible case. Specifically there
exists also the solution in which
\begin{equation}
\phi\left(  t\right)  =\phi_{0}+\frac{2p}{1+3p}K_{0}t^{1+3p}+\frac{2p}%
{1+3p}\ln\left(  \left(  1+3p\right)  t\right)  , \label{fttb.48}%
\end{equation}
where now the potential as a function of $t$ is expressed as
\begin{equation}
V\left(  t\right)  =\frac{\left(  3p-1\right)  }{\left(  3p+1\right)  }%
\frac{6p^{2}}{t^{2}}-12p^{2}K_{0}t^{-1+3p}. \label{fttb.49}%
\end{equation}
From the latter expressions we observe that solution (\ref{fttb.40}) is
recovered when $K_{0}=0$, while $K_{0}$ is nothing else than a constant of
integration. \ There are two constants of integration as equation
(\ref{ftb.19}) is a second-order differential equation in terms of $\phi$.
Furthermore from (\ref{fttb.48}) we find that the scalar-field potential is
given in terms of the Lambert $W\left(  \phi\right)  $ Function. \ In the
following we consider the case in which $K_{0}=0$.

In order to study the stability of the power-law solution for the theory
(\ref{fttb.42}) as in the case of the de\ Sitter Universe we prefer to work
with equation (\ref{fttb.28}). Hence, if we substitute~$~H\left(  t\right)
=\frac{p}{t}+\varepsilon H_{P}\left(  t\right)  ~$into (\ref{fttb.28}) with
(\ref{fttb.42}) and we linearize around $\varepsilon=0$, the following linear
equation is found,%
\begin{equation}
\ddot{H}_{P}+3t\left(  1+p\right)  \dot{H}+6pH_{P}=0 \label{fttb.43}%
\end{equation}
with closed-form solution%
\begin{equation}
H_{P}\left(  t\right)  =\frac{p}{t}\left(  H_{P}^{1}t^{-1}+H_{P}^{1}%
t^{1-3p}\right)  , \label{fttb.44}%
\end{equation}
which means that, as $t$ increases, $H_{P}\left(  t\right)  $ decays,~$H_{P}%
\rightarrow0$ and the power-law solution is stable for every $p>0$.

We now study the stability of the dust solution, $a\left(  t\right)
=a_{0}t^{\frac{2}{3}}$, in the present of the cosmological constant, that is,
of the theory%
\begin{equation}
\Phi\left(  B\right)  =-\frac{2}{9}B\ln\left(  B\right)  -2\Lambda.
\label{fttb.45}%
\end{equation}

\begin{figure}[ptb]
\includegraphics[height=8cm,width=8.5cm]{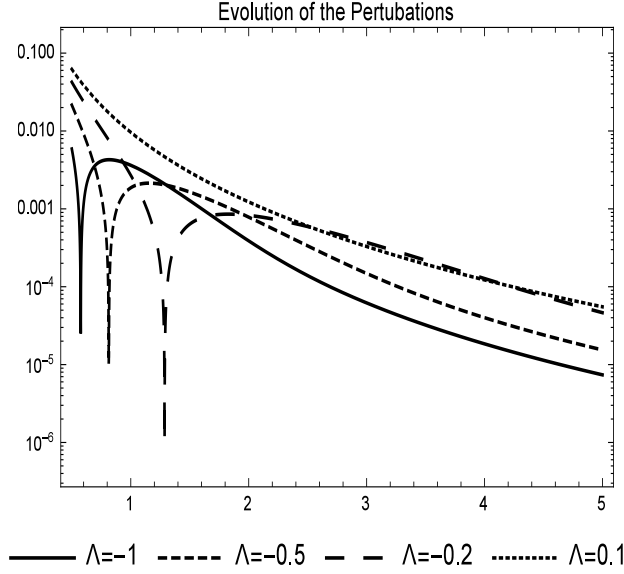}\centering
\caption{Evolution of the perturbations $\left\vert H_{P}\left(  t\right)
\right\vert $ given by the expression (\ref{fttb.47}) with $H_{P}^{0}=0$ and
$H_{P}^{1}=10^{-2}$ for various values of the cosmological constant $\Lambda
.$}%
\label{plot2}%
\end{figure}

Therefore from (\ref{fttb.28}) we find the linear equation%
\begin{equation}
4t^{2}\ddot{H}_{P}+t\left(  20-9\Lambda t^{2}\right)  \dot{H}_{P}+4\left(
4-9\Lambda^{2}t^{2}\right)  H_{P}=0 \label{fttb.46}%
\end{equation}
with closed-form solution,%
\begin{equation}
H_{P}\left(  t\right)  =t^{-2}\exp\left(  \frac{9}{8}\Lambda t^{2}\right)
\left(  H_{P}^{0}+H_{P}^{1}\int e^{-\frac{9}{8}\Lambda t^{2}}\exp\left(
-e^{-\frac{9}{8}\Lambda t^{2}}\right)  \right)  dt, \label{fttb.47}%
\end{equation}
from which we observe that $H_{P}\left(  t\right)  $ decays for every negative
value of $\Lambda${, while for $\Lambda>0$ the perturbations are dominant and
the dust solution becomes unstable, except if the initial conditions are such
that $H_{P}^{0}=0$. In Fig. \ref{plot2} we give the evolution of the
perturbations for $H_{P}^{0}=0$ and for various values of the parameter
$\Lambda,$ from which we observe that the perturbations decay. }

\section{Stability of scaling solutions in the presence of matter}

\label{scaling}

In the previous Section we studied the case in which the geometric dark energy
fluid has a constant equation of state parameter in the case of vacuum.\ In
this Section we consider also the existence of a matter source with constant
equation of state parameter and we study the stability of scaling solutions.

Let $w_{m}=\gamma-1$ be the equation of state parameter for the matter source,
that is, $p_{m}=\left(  \gamma-1\right)  \rho_{m},$ and assume that the
universe is dominated by the matter. Hence the scale factor is $a\left(
t\right)  =t^{\frac{2}{3\gamma}}$ while from equation (\ref{ftb.20a}) it
follows that $\rho_{m}=\rho_{m0}a^{-3\gamma}$. \ There are two possibilities
for the geometric dark energy in order that $a\left(  t\right)  =t^{\frac
{2}{3\gamma}}$ be an exact solution. It is possible that the scalar field
mimics the matter source which, as we see below, is similar to the study of
the previous Section or, alternatively, for the dark energy fluid to be
canceled without the scalar field to be zero or constant.

\subsection{Scalar field mimics the matter source}

Consider that the scalar field mimics the matter source. Then for the scale
factor, $a\left(  t\right)  =a_{0}t^{\frac{2}{3\gamma}}$, from equations
(\ref{ftb.18})-(\ref{ftb.20}) it follows
\begin{equation}
\phi\left(  t\right)  =\phi_{0}+\frac{\gamma}{2+\gamma}K_{0}t^{\frac{2+\gamma
}{\gamma}}+\frac{\left(  4-3a^{-3\gamma}\gamma^{2}\rho_{m0}\right)  }{3\left(
2+\gamma\right)  }\ln t \label{fttb.50}%
\end{equation}
and%
\begin{equation}
V\left(  t\right)  =\frac{2a_{0}^{-3\gamma}\left(  2-\gamma\right)  \left(
4-3a^{-3\gamma}\gamma^{2}\rho_{m0}\right)  }{3\gamma^{2}\left(  2+\gamma
\right)  }t^{-2}-\frac{4}{\gamma}K_{0}t^{-2+\frac{2+\gamma}{\gamma}},
\label{fttb.51}%
\end{equation}
where the scalar field potential, $V\left(  \phi\right)  $, is expressed in
terms of the Lambert $W\left(  \phi\right)  $ Function or, when $K_{0}=0$, as
the exponential potential%
\begin{equation}
V\left(  \phi\right)  =\frac{2a_{0}^{-3\gamma}\left(  2-\gamma\right)  \left(
4-3\gamma^{2}\rho_{m0}\right)  }{3\gamma^{2}\left(  2+\gamma\right)  }%
\exp\left(  -\frac{6\left(  2+\gamma\right)  }{\left(  4-3\gamma^{2}\rho
_{m0}\right)  }\left(  \phi-\bar{\phi}_{0}\right)  \right)  . \label{fttb.52}%
\end{equation}

Furthermore for the latter potential the $\Phi\left(  B\right)  $ is given by
the expression $\Phi\left(  B\right)  =\frac{8-3\gamma^{2}\rho_{m0}}{12\left(
2+\gamma\right)  }B\ln B$.

As far the stability of that latter solution is concerned, we perform a
perturbation of the form $a\left(  t\right)  =a_{0}t^{\frac{2}{3\gamma}}%
+a_{0}t^{\frac{2}{3\gamma}}a_{\varepsilon}\left(  t\right)  $ and we find the
third-order differential equation%
\begin{align}
0  &  =\left(  8-3\gamma^{2}\rho_{m0}\right)  \left(  \gamma t^{3}%
a_{\varepsilon}^{\left(  3\right)  }+\left(  2+3\gamma\right)  t^{2}\ddot
{a}_{\varepsilon}\right)  +\nonumber\\
&  +2\left(  16+3\gamma\left(  \left(  2+\gamma\right)  \gamma-4\right)
\rho_{m0}\right)  \dot{a}_{\varepsilon}-\left(  6\gamma\left(  \gamma
^{2}-4\right)  \rho_{m0}\right)  a_{\varepsilon} \label{fttb.53}%
\end{align}
which under the Kummer transformation, $t->e^{\tau}$, becomes the autonomous
equation,%
\begin{align}
0  &  =\left(  8-3\gamma^{2}\rho_{m0}\right)  \left(  \gamma a_{\varepsilon
}^{\left(  3\right)  }\left(  \tau\right)  +2a_{\varepsilon}^{\left(
2\right)  }\left(  \tau\right)  \right)  +\nonumber\\
&  -\left(  2-\gamma\right)  \left(  8+3\gamma\left(  4+\gamma\right)
\rho_{m0}\right)  a_{\varepsilon}^{\left(  1\right)  }\left(  \tau\right)
+6\gamma\left(  \gamma^{2}-4\right)  \rho_{m0}a_{\varepsilon}\left(
\tau\right)  , \label{fttb.54}%
\end{align}
which admits the general solution,
\begin{equation}
a_{\varepsilon}\left(  \tau\right)  =a_{p}^{1}\exp\left(  \lambda_{+}%
\tau\right)  +a_{p}^{2}\exp\left(  \lambda_{-}\tau\right)  +a_{p}^{3}\left(
e^{-x}\right)  . \label{fttb.55}%
\end{equation}
in which%
\begin{equation}
\lambda_{\pm}=-\frac{2+\gamma}{2\gamma}\pm\sqrt{\frac{\left(  2-\gamma\right)
\left(  8\left(  2-\gamma\right)  -3\gamma^{2}\left(  18+7\gamma\right)
\rho_{m0}\right)  }{4\gamma^{2}\left(  8-3\gamma^{2}\rho_{m0}\right)  }}.
\label{fttb.56}%
\end{equation}

The perturbations decay when the real parts of\ $\lambda_{\pm}$ are negative,
that is, $\operatorname{Re}\left(  \lambda_{+}\right)  <0$ and
$\operatorname{Re}\left(  \lambda_{-}\right)  <0$. Hence we find that for
$\gamma\in\lbrack1,2)$
\begin{equation}
\rho_{m0}\leq\frac{8\left(  2-\gamma\right)  }{3\gamma^{2}\left(
18+7\gamma\right)  }. \label{fttb.57}%
\end{equation}

\subsection{Dark energy fluid is canceled}

As we discussed above, there is also the alternate scenario in which the dark
energy fluid is canceled and does not contribute to the solution, without the
scalar field to be zero or constant. That is an analogue of the analysis in
\cite{Ratra,Liddle11,robert,gong} for the canonical scalar field and others
\cite{take1,take2,take3,take4,take5,take6}.

We substitute the power-law solution, $a\left(  t\right)  =a_{0}t^{\frac{2}%
{3}\gamma}$, with $\gamma\in\lbrack1,2)~$into the field equations
(\ref{ftb.18})-(\ref{ftb.20}) and we find that the geometric dark fluid is
canceled when $\rho_{m0}=\frac{4}{3\gamma^{2}}a_{0}^{3\gamma}$, and
\begin{equation}
\phi\left(  t\right)  =\phi_{0}t^{\lambda}~,~V\left(  \phi\right)  =V_{0}%
\phi^{\mu}, \label{fttb.58}%
\end{equation}
where%
\begin{equation}
\mu=\frac{2-\gamma}{2+\gamma},~\lambda=\frac{2+\gamma}{\gamma}\text{ and
}V_{0}=-\frac{4\left(  2+\gamma\right)  }{\gamma^{2}}\left(  \phi_{0}\right)
^{\frac{2\gamma}{2+\gamma}}\,.~ \label{fttb.59}%
\end{equation}

As far as the $\Phi\left(  B\right)  $ function is concerned, it is
straightforward to see that for the power-law potential the corresponding
theory is power-law, that is, $\Phi\left(  B\right)  =\Phi_{0}B^{\frac{\mu
}{\mu-1}}$.

The differential equation which drives the evolution of the scalar field is
equation (\ref{ftb.19}), which for the power-law solution becomes%
\begin{equation}
\ddot{\phi}+\frac{V_{0}}{2}\phi^{\mu}=0. \label{fttb.60}%
\end{equation}
This equation has a movable singularity\footnote{For every $\mu$ no positive
integer.} and in order to remove it we perform the change of variable
$\phi\left(  t\right)  =\phi_{0}t^{\frac{2+\gamma}{\gamma}}\left(  \psi\left(
t\right)  \right)  ,~$while we apply the Kummer transformation $t=e^{\tau}$ to
write the differential equation (\ref{fttb.60}) as the second-order autonomous
equation%
\begin{equation}
\gamma^{2}\psi^{\left(  2\right)  }\left(  \tau\right)  +\left(
4+\gamma\right)  \gamma\psi^{\left(  1\right)  }\left(  \tau\right)  +2\left(
2+\gamma\right)  \left(  1-\left(  \psi\left(  \tau\right)  \right)
^{-\frac{2\gamma}{2+\gamma}}\right)  \psi\left(  \tau\right)  =0
\label{fttb.61}%
\end{equation}
or, equivalently,%
\begin{align}
\dot{\psi}  &  =p_{\psi},\label{fttb.62}\\
\dot{p}_{\psi}  &  =-\frac{\left(  4+\gamma\right)  }{\gamma}p_{\psi}%
-\frac{2\left(  2+\gamma\right)  }{\gamma^{2}}\left(  1-\left(  \psi\left(
\tau\right)  \right)  ^{-\frac{2\gamma}{2+\gamma}}\right)  \psi\left(
\tau\right)  , \label{fttb.63}%
\end{align}
which in the range $\gamma\in\lbrack1,2)$ admits the critical point
$P_{1}=\left(  1,0\right)  .$ Point $P_{1}$ describes the solution
(\ref{fttb.59}).

We linearize the system, (\ref{fttb.62}), (\ref{fttb.63}), around the critical
point, $P_{1}$, and the eigenvalues of the linearized system are $e_{1}\left(
P_{1}\right)  =-1$ and $e_{2}\left(  P_{1}\right)  =-\frac{4}{\gamma}$, which
means that solution (\ref{fttb.59}) is stable.

\subsection{Leading-order behaviour}

Assume now that the matter source, $\rho_{m}$, dominates and in contrast to
the above we assume that $a\left(  t\right)  \simeq t^{\frac{2}{3\gamma}}$ is
the leading-order behaviour of the scale factor. We now study the evolution of
the scalar field by studying the dynamics of equation (\ref{ftb.19}).

When we substitute $H\left(  t\right)  =\frac{2}{3\gamma t}$ in (\ref{ftb.19})
with the power-law potential, $V\left(  \phi\right)  =V_{0}\phi^{\mu}$, we
find that%
\begin{equation}
\ddot{\phi}+V_{0}\phi^{\mu}-2\left(  1-\gamma\right)  \left(  4-3\gamma
^{2}\rho_{m0}\right)  =0, \label{fttb.64}%
\end{equation}
where, because we have assumed that the matter source dominates, it follows
that $\left(  4-3\gamma^{2}\rho_{m0}\right)  \simeq0$ and $\mu>0$. Hence the
latter equation takes the form of (\ref{fttb.60}) in which now $\mu$ is not
related to $\gamma$ as before. Equation (\ref{fttb.66}) has the closed-form
solution $\phi\left(  t\right)  =\phi_{0}t^{\frac{2}{1-\mu}}$ with
$V_{0}=-\frac{4\left(  1+\mu\right)  }{\left(  1-\mu\right)  ^{2}}\phi
_{0}^{1-\mu}$, from where we have that, if $\mu=\frac{2-\gamma}{2+\gamma}$,
then solution (\ref{fttb.59}) recovered. The theory with $\mu=1$ is not of
special interest because that corresponds to the linear $\Phi\left(  B\right)
=B$ theory, while we have assumed that $\Phi_{,BB}\neq0$.

We apply the transformation $\phi\left(  t\right)  =\phi_{0}t^{\frac{2}{1-\mu
}}\left(  \psi\left(  t\right)  \right)  ,~t=e^{\tau}$, where we find the
autonomous equation%
\begin{equation}
\left(  1-\mu\right)  ^{2}\psi^{\left(  2\right)  }\left(  \tau\right)
-\left(  \mu^{2}+2\mu-3\right)  \psi^{\left(  1\right)  }\left(  \tau\right)
+2\left(  1+\mu\right)  \left(  1-\left(  \psi\left(  \tau\right)  \right)
^{\mu-1}\right)  \psi\left(  \tau\right)  =0. \label{fttb.65}%
\end{equation}

The latter can be written as
\begin{align}
\dot{\psi}  &  =p_{\psi},\label{fttb.66}\\
\dot{p}_{\psi}  &  =\left(  \mu^{2}+2\mu-3\right)  p_{\psi}-2\left(
1+\mu\right)  \left(  1-\left(  \psi\left(  \tau\right)  \right)  ^{\mu
-1}\right)  \psi\left(  \tau\right)  . \label{fttb.67}%
\end{align}
\newline

For arbitrary power $\mu$ the latter system admits as critical point only the
point $P_{1}$, while when $\left(  \mu-1\right)  $ is an even integer the
system admits the extra additional points $P_{2}=\left(  -1,0\right)  $ and
$P_{0}=\left(  0,0\right)  $. Note that $P_{0}~$is a critical point of the
system for any $\mu>1.$

\begin{figure}[ptb]
\includegraphics[height=12.5cm]{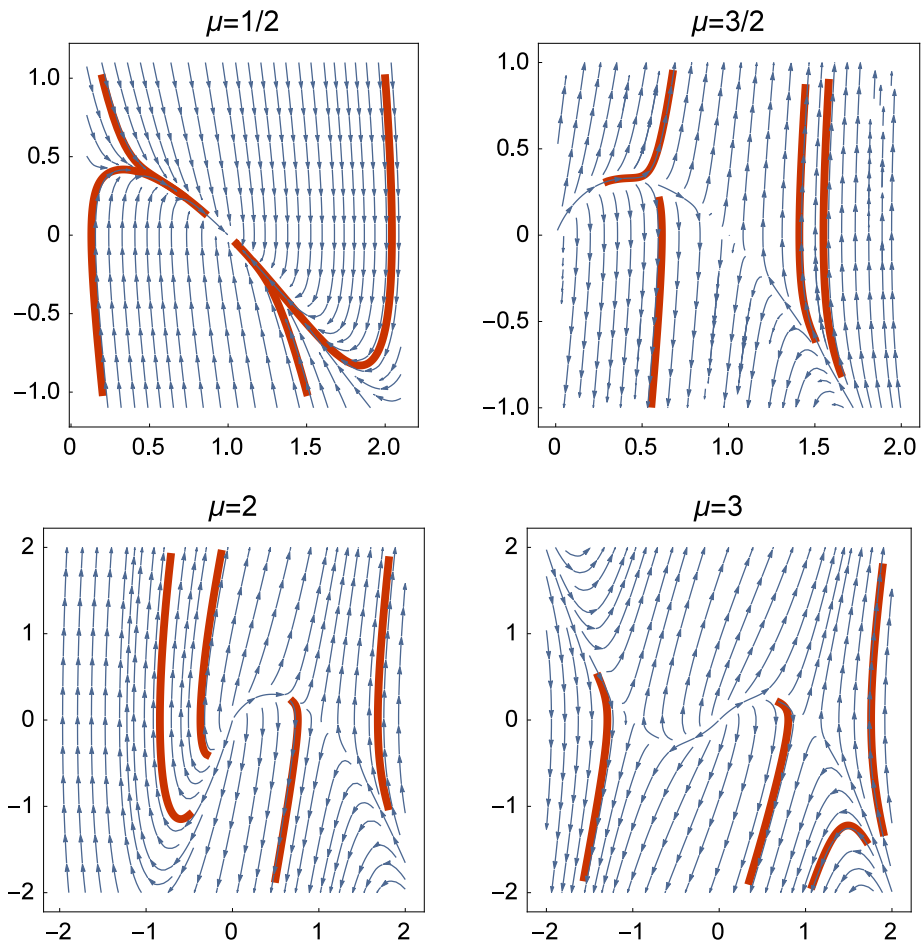}\centering
\caption{Phase portrait for the dynamical system (\ref{fttb.66}),
(\ref{fttb.67}) for various values of the free parameter $\mu$. The plots are
for $\mu=\frac{1}{2}$, where the critical point $P_{1}$ is an attractor, for
$\mu=\frac{3}{2}$ where $P_{1}$ is a hyperbolic point, for $\mu=2$ where the
two critical points $P_{1}$ and $P_{0}$ are hyperbolic and finally for $\mu=3$
in which all the critical points, $P_{1},~P_{2}$ and $P_{0}$ are unstable.
Solid lines describe solutions of the differential equations.}%
\label{plot4}%
\end{figure}

Easily from the eigenvalues of the linearized system around the critical
points we find that $P_{1}$ is a stable point for every $\mu<1$, while $P_{0}$
and $P_{2}$, when they exist, are hyperbolic points. In Fig. \ref{plot4} the
phase-space diagram of the dynamical system is presented for various values of
the parameter $\mu$. \ Note that as above $P_{1}$ as a stable point means that
the scaling solution is a stable solution, and actually in the limit in which
$\mu=\frac{2-\gamma}{2+\gamma}$, \ for $\gamma\in\lbrack1,2)$ it follows that
$\mu<1$ and the analysis above is recovered.

Finally the energy density for the scalar field is calculated to be
$\rho_{\phi}\simeq\rho_{m0}t^{\frac{2\mu}{1-\mu}},~\rho_{m0}=\rho_{m0}\left(
\mu,\gamma,\phi_{0}\right)  $. It is important to mention here that from
(\ref{fttb.60}) we find that $p_{\phi}\simeq0$. However, it is a weak
equivalence in the sense that the coefficient of the term, $p_{\phi}$, for the
leading-order behaviour that we consider is zero, which means that very close
to the singularity the scalar field acts like a dust fluid. That is a
different result from the same analysis for a minimally coupled scalar field
where it was found that the scalar field can mimic an ideal gas with arbitrary
equation of state parameter \cite{sahni1,Liddle11}.

\subsubsection{Negative power}

In order to complete our study with the scaling solution we assume that the
power, $\mu$, is negative. Hence the singular solution, $\phi\left(  t\right)
=\phi_{0}t^{\frac{2}{1-\mu}},~$provides that the in the matter-dominated era
the scalar field dominates the universe.

Now for negative values of $\mu$ the dynamical system (\ref{fttb.66}),
(\ref{fttb.67}) admits only the critical points, $P_{1}$ and $P_{2}$, when
$\left\vert \mu-1\right\vert $ is an even number. As above we calculate the
eigenvalues of the linearized system and we find that for the point, $P_{1}%
$,~$e_{1}\left(  P_{1}\right)  <0$, $e_{2}\left(  P_{2}\right)  <0$ when
$\left\vert \mu\right\vert <1$, while for $\mu<-1$, $P_{1}$ is a saddle point.
Furthermore, as far as the point, $P_{2}$, is concerned, we find that one of
the eigenvalues is always positive which means that $P_{2}$ is a saddle point.

\begin{figure}[ptb]
\includegraphics[height=12.5cm]{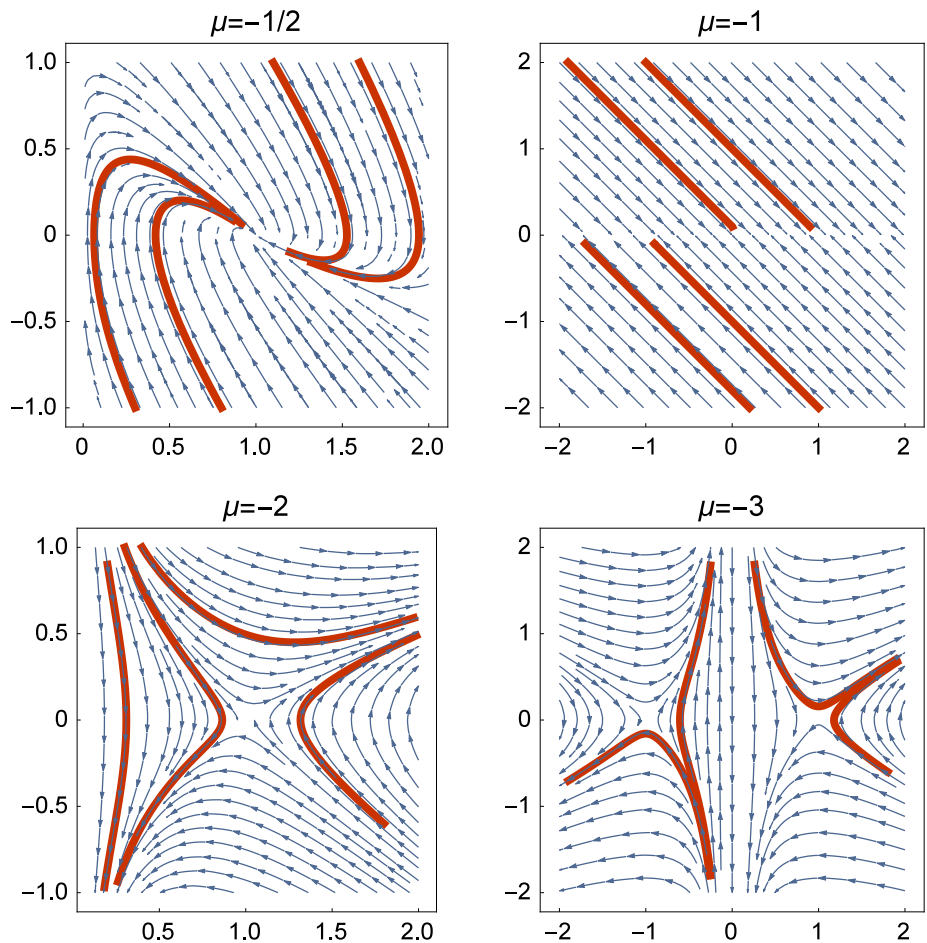}\centering
\caption{Phase portrait for the dynamical system (\ref{fttb.66}),
(\ref{fttb.67}) for negative values of the potential $\mu$. The plots are for
$\mu=-\frac{1}{2},~\mu=-1$,~$\mu=-2$ and $\mu=-3$ in order demonstrate all the
possible cases of stability. Solid lines describe solutions of the
differential equations.}%
\label{plot5}%
\end{figure}

However, for the special value, $\mu=-1$, the differential equation
(\ref{fttb.65}) becomes%
\begin{equation}
\psi^{\left(  2\right)  }\left(  \tau\right)  +\psi^{\left(  1\right)
}\left(  \tau\right)  =0 \label{fttb.68}%
\end{equation}
which gives the solution, $\psi\left(  \tau\right)  =\psi_{0}e^{-\tau}%
+\psi_{1},$ that is, $\psi\left(  t\rightarrow+\infty\right)  \simeq\psi_{1}$.
The points $P_{1},$ $P_{2}$ are just two points on the solution at the limit
$\psi\left(  \tau\right)  =\psi_{1}$. In Fig. \ref{plot5} the phase portrait
of the system (\ref{fttb.66}), (\ref{fttb.67}) is presented for negative
values of the power, $\mu$, in order to demonstrate all the possible cases.

There is an important difference with the canonical scalar field and that is
that there are no oscillatory models. The critical points always have real
eigenvalues which means that close to the critical points the scalar field is
not oscillating.

\section{Algebraic Solution}

\label{algebraic}

There are various ways in which the precise meaning for the solution of a
system of differential equations can be cast. Usually, when we refer to the
solution of a differential equation, we mean that there exists a set of
explicit functions describing the variation of the dependent variables with
the independent variable. These solutions are called closed-form solutions and
the exact solutions we presented before are a special class of this kind of
solutions. Another way to cast the solution of a differential equation is to
describe the differential equation as system of independent\ first integrals
and invariants, while usually in the latter scenario the equivalent system can
be written as system of algebraic equations. That expression is called
algebraic solution of the differential equation.

Indeed these are different ways to describe the solution of a differential
equation and yet there exists a central feature amongst them. The above
different descriptions are directly related with the existence of
transformations in which the system and the solution are invariant, that is
the existence of symmetries.

\subsection{Integrability}

We assume that the matter source, $\rho_{m},~p_{m}$, admits an equation of
state parameter, $w_{m}\left(  a\right)  $, such that the Lagrangian term is
$L_{m}=N\rho_{m0}\exp\left(  -3w_{m}\left(  a\right)  \ln\left(  a\right)
\right)  $. Hence, from the minisuperspace description of the field equations,
from Lagrangian, (\ref{ftb.16}), we observe that the gravitational field
equations can be seen as the equations of motion for a particle moving on a
two-dimensional Riemannian manifold, the minisuperspace, under the action of
an effective potential. Moreover the lapse function, $N$, is the one which
provides the constraint equation, (\ref{ftb.18}). \ Hence techniques from
Classical Mechanics can be applied for the determination of conservation laws
and the derivation of solutions. Indeed various approaches have been applied
in the literature \cite{angrg,take2,noref1,noref2,noref3,noref4}.

It is well known that there exists a unique relation between the symmetries of
this kind of systems of differential equations with the symmetries which
define the underlying geometry \cite{noref5,noref6}. That means that any
generator of a symmetry vector for the dynamical system has to be a symmetry
also for the geometry. For instance the conservation law of momentum for the
free particle follows from the translation symmetry of the Euclidean
spacetime. The group of translations with the group of rotations form the
group of isometries or Killing vectors of the Euclidean space.

By definition a Killing vector in a Riemannian manifold is the generator of
the transformation which keep invariant the length and the angles. On the
other hand, a Homothetic vector is the generator of the transformation which
keep invariant the angles and rescale by a constant the length, whereas a
Conformal vector is called the generator of the transformation which preserves
the angles on the space.

Now \ for autonomous Hamiltonian systems the \textquotedblleft
Energy\textquotedblright\ denotes the volume in the phase space. For any
isometry which leave invariant this volume in the phase space corresponds a
conservation law which commutes with the Hamiltonian. As far as concerns the
Homothetic vector, the solutions can be transformed under other solution but
with a rescaled \textquotedblleft Energy\textquotedblright\ value. These two
transformations relates objects which are congruent, with the identical
congruent to be provided by the isometries.

The situation is totally different under conformal transformations. Indeed
Hamiltonian systems are not invariant under conformal transformations except
if the \textquotedblleft Energy\textquotedblright\ is zero \cite{angrg}, which
means that the volume in the phase space has dimensions zero. Moreover the
volume continues to be zero under conformal transformations and consequently
conservation laws can be constructed.

In order to demonstrate that mathematically, consider $\mathcal{H}\left(
\mathbf{p,q}\right)  =0$ to be the energy of an autonomous Hamiltonian system
and $I\left(  \mathbf{p,q}\right)  $ be a conservation law generated by a
conformal vector. Then it follows that there exists a function, $\omega$, such
that $D_{t}\left(  I\right)  =I_{,t}+\left\{  I,\mathcal{H}\right\}
=\omega\mathcal{H}$; that is, $D_{t}\left(  I\right)  =0$, which means that
$I$ is a conservation law. These kinds of conservation laws are generated by
nonlocal symmetries, which reduce to local when $\omega=const$ or $\omega=0$.

Because of the constraint equation, (\ref{ftb.18}), we can say that the Energy
of the Mechanical analogue is zero and construct conservation laws by using
the conformal algebra of the minisuperspace. Indeed, as has been shown in
\cite{chris1}, for every Conformal vector field there corresponds a
conservation law for the field equations, for any function, $V\left(
\phi\right)  $. Moreover, because the minisuperspace has dimension two, it
admits an infinite-dimensional conformal algebra, that is, there exists an
infinitenumber of (nonlocal) conservation laws. Of course these conservation
laws are not in involution with each other, but they are with the Hamiltonian
applying the constraint equation,~$\mathcal{H}\left(  \mathbf{p,q}\right)  =0$.

Furthermore the degrees of freedom of the field equations are two and while
the constraint equation can be seen as a conservation law and because there
exists at least one function which is in ivolution with $\mathcal{H}$,\ we can
say that the gravitational field equations followed by the Lagrangian
(\ref{ftb.16}) are integrable for an arbitrary function, $V\left(
\phi\right)  $. That approach was applied recently in \cite{ssol1,ssol2} to
construct the solution of the canonical minimally coupled scalar field.

For simplicity, in the following we assume that the spacetime is empty.
However, in a similar way the algebraic solution can be constructed for any
matter source.

\subsection{Solution in terms of the scale factor}

For a general lapse function $N\left(  t\right)  $ the gravitational field
equations (\ref{ftb.18})-(\ref{ftb.20}) are%
\begin{equation}
3\left(  \frac{\dot{a}}{aN}\right)  ^{2}=\left(  \frac{\dot{a}}{aN}\right)
\frac{\dot{\phi}}{N}+\frac{1}{2}V, \label{fttb.69}%
\end{equation}%
\begin{equation}
2\left(  \frac{\ddot{a}}{aN^{2}}-\frac{\dot{a}\dot{N}}{aN^{3}}\right)
+\left(  \frac{\dot{a}}{aN}\right)  ^{2}=\frac{\ddot{\phi}}{N^{2}}-\frac
{\dot{\phi}\dot{N}}{N^{3}}+\frac{1}{2}V \label{fttb.70}%
\end{equation}
and%
\begin{equation}
\frac{1}{6}V_{,\phi}+\left(  \frac{\ddot{a}}{aN^{2}}-\frac{\dot{a}\dot{N}%
}{aN^{3}}\right)  +\left(  \frac{\dot{a}}{aN}\right)  ^{2}=0 \label{fttb.71}%
\end{equation}
from where we can see that the dark energy equation of state parameter is%
\begin{equation}
w_{\phi}=-\frac{\ddot{\phi}-\dot{\phi}\left(  \ln N\right)  ^{\cdot}+\frac
{1}{2}N^{2}V}{3\frac{\dot{a}}{a}\dot{\phi}+\frac{1}{2}N^{2}V}. \label{fttb.72}%
\end{equation}

Without loss of generality we can consider locally for any solution there
exists a lapse function such that $a\left(  t\right)  =e^{t}$. In the
following we replace $t$ with $\tau$ in order to make clear that we are in the
frame in which $\tau=\ln a$.

Therefore equations (\ref{fttb.69}) and (\ref{fttb.70}) become%
\begin{equation}
\frac{d\phi}{d\tau}+\frac{1}{6}N\left(  a\right)  ^{2}V\left(  \phi\left(
a\right)  \right)  -1=0 \label{fttb.73}%
\end{equation}
and%
\begin{equation}
N\frac{d^{2}\phi}{d\tau^{2}}-\frac{d\phi}{d\tau}\frac{dN}{d\tau}+\frac{1}%
{2}N^{3}V+\frac{dN}{d\tau}=0. \label{fttb.74}%
\end{equation}

From (\ref{fttb.73}) we substitute for the lapse function $N\left(  a\right)
$ and find
\begin{equation}
N^{2}\left(  \tau\right)  =6\left(  1-\frac{d\phi}{d\tau}\right)  \left(
V\left(  \phi\right)  \right)  ^{-1}. \label{fttb.75}%
\end{equation}
Hence expression (\ref{fttb.74}) becomes%
\begin{equation}
\left(  \frac{d^{2}\phi}{d\tau^{2}}+6\left(  1-\frac{d\phi}{d\tau}\right)
\right)  +\left(  \ln V\left(  \phi\right)  \right)  _{,\phi}\left(
2-\frac{d\phi}{d\tau}\right)  \left(  1-\frac{d\phi}{d\tau}\right)  =0.
\label{fttb.76}%
\end{equation}
This expression is an autonomous differential equation and easily solved if we
define the new variable$~\frac{d\phi}{da}=\Psi\left(  \phi\right)  $. It
becomes the first-order equation
\begin{equation}
\left(  \Psi\frac{d\Psi}{d\phi}+6\left(  1-\Psi\right)  \right)  +\left(  \ln
V\left(  \phi\right)  \right)  _{,\phi}\left(  2-\Psi\right)  \left(
1-\Psi\right)  =0, \label{fttb.77}%
\end{equation}
which in general is an Abel's equation. Furthermore the lapse function
(\ref{fttb.75}) can be written as a function of $\phi$ as follows%
\begin{equation}
N^{2}\left(  \phi\right)  =6\left(  1-\Psi\left(  \phi\right)  \right)
\left(  V\left(  \phi\right)  \right)  ^{-1}. \label{fttb.77a}%
\end{equation}

In order to demonstrate our result we consider the simplest form for the
potential that admits a stable de Sitter solution, that is, $V\left(
\phi\right)  =V_{0}e^{-3\phi}$. Hence from (\ref{fttb.77}) we find that
$\Psi\left(  \phi\right)  =1-\Psi_{0}e^{3\phi}$ or%
\begin{equation}
\phi\left(  \tau\right)  =-\frac{1}{3}\ln\left(  1+\Psi_{0}e^{3\left(
\tau-\tau_{0}\right)  }\right)  +\left(  \tau-\tau_{0}\right)  .
\label{fttb.78}%
\end{equation}

For $\Psi_{0}\neq0~$from (\ref{fttb.75}) we calculate the lapse
function$~N^{2}\left(  \tau\right)  =\frac{V_{0}}{6\Psi_{0}}\left(  \Psi
_{0}+e^{-3\tau}\right)  ^{-2}$. Hence the Hubble function $H\left(  a\right)
=\frac{1}{N}\frac{\dot{a}}{a}$ is calculated to be%
\begin{equation}
\left(  \frac{H\left(  a\right)  }{H_{0}}\right)  ^{2}=\left(  \alpha_{1}%
\Psi_{0}+\alpha_{2}\left(  \frac{a}{a_{0}}\right)  ^{-3}\right)  ^{2}
\label{fttb.79}%
\end{equation}
in which $\alpha_{1}=\sqrt{\frac{\Psi_{0}V_{0}}{6}}H_{0}$ and $\alpha
_{2}=\sqrt{\frac{V_{0}}{6\Psi_{0}}}H_{0}$.

\subsubsection{Solution for arbitrary potential}

Without loss of generality we replace in (\ref{fttb.76}) $V\left(
\phi\right)  =\exp\left(  \int\lambda\left(  \tau\right)  d\tau\right)  $ and
$\frac{d\phi}{d\tau}=\sigma\left(  \tau\right)  $, that is,
\begin{equation}
\frac{\sigma}{1-\sigma}\frac{d\sigma}{d\tau}+6\sigma+\left(  2-\sigma\right)
\lambda=0. \label{fttb.80}%
\end{equation}

Hence for a known function, $\sigma\left(  \tau\right)  $, we have the
algebraic equation
\begin{equation}
\lambda=\Sigma\left(  \sigma,\frac{d\sigma}{d\tau}\right)  , \label{fttb.81}%
\end{equation}
that is, $V\left(  \phi\right)  =\exp\left(  \int\Sigma\left(  \sigma
,\frac{d\sigma}{d\tau}\right)  d\tau\right)  .$ Therefore all the functions
have been expressed in terms of the scale factor. Finally the lapse, $N\left(
\tau\right)  $, is calculated to be%
\begin{equation}
N^{2}\left(  \tau\right)  =6\frac{\left(  1-\sigma\left(  \tau\right)
\right)  }{\sigma\left(  \tau\right)  }\exp\left(  -\int\Sigma\left(
\tau\right)  d\tau\right)  . \label{fttb.82}%
\end{equation}

The question that can be raised is why that specific lapse and not another
one. Mathematically we saw the final equation is autonomous which easily is
reduced to a first-order differential equation. On the other hand, physically
on that lapse the Hubble function, $H\left(  \ln\left(  a\right)  \right)
=\left(  N\left(  \ln\left(  a\right)  \right)  \right)  ^{-1}$,~as also all
the physical quantities are expressed directly in terms of the scale factor.

Moreover, the reason that the final solution is expressed in terms of an
arbitrary function, say $\Sigma$, is directly related with the general
functional form for potential that we have assumed. There are various ways to
constrain the potential, as for instance to set a specific equation of state
parameter for the dark energy fluid. An approach that we followed to constrain
the potential in the case of a canonical scalar field \cite{anprd11}.

\subsection{Reproduce the $\Lambda$-Cosmology}

As a nontrivial example consider that the lapse function is
\begin{equation}
N\left(  \tau\right)  ^{-2}=\Omega_{\Lambda}+\Omega_{m0}e^{-3\tau},
\label{fttb.83}%
\end{equation}
where the coresponding Hubble function is that of the $\Lambda$CDM model. We
search for the potential such that the scalar field mimics the $\Lambda$CDM model.

We find the solution%
\begin{align}
\phi\left(  \tau\right)   &  =\phi_{0}+\frac{2}{3}\left(  \tau-\frac
{2\Omega_{\Lambda}}{3\Omega_{m0}}e^{3\tau}\right)  +\phi_{1}\frac{\sqrt
{\Omega_{\Lambda}e^{6\tau}+\Omega_{m0}e^{3\tau}}}{3\Omega_{\Lambda}%
}\nonumber\\
&  -\phi_{1}\frac{\Omega_{m0}}{\left(  \Omega_{\Lambda}\right)  ^{\frac{3}{2}%
}}\ln\left(  \Omega_{\Lambda}e^{\frac{3}{2}\tau}+\sqrt{\left(  \Omega
_{\Lambda}\right)  ^{2}e^{3\tau}+\Omega_{m0}}\right)  , \label{fttb.84}%
\end{align}
and potential%
\begin{equation}
V\left(  \tau\right)  =10\Omega_{\Lambda}+8\frac{\left(  \Omega_{\Lambda
}\right)  ^{2}}{\Omega_{m0}}e^{3\tau}+2e^{-3\tau}\Omega_{m0}-6\phi_{1}%
\sqrt{\Omega_{\Lambda}e^{6\tau}+\Omega_{m0}e^{3\tau}}. \label{fttb.85}%
\end{equation}
The constants, $\phi_{0}~$and $\phi_{1}$, are constants of integration.

\begin{figure}[ptb]
\includegraphics[height=7cm]{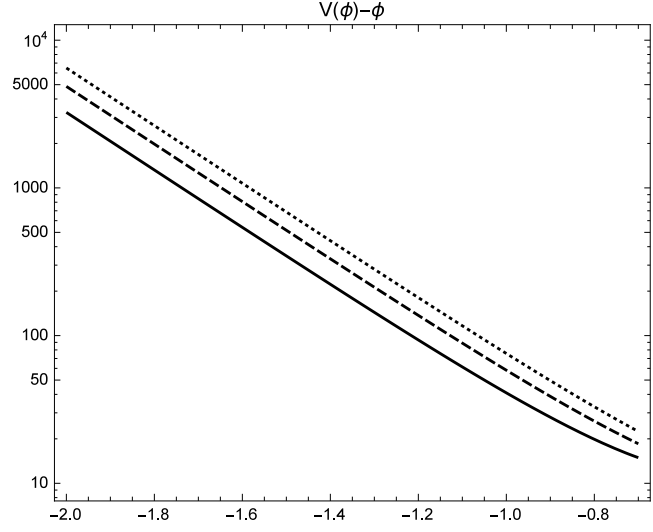}\centering
\caption{Qualitative behaviour of the potential (\ref{fttb.87}) for different
values of the parameters $\Omega_{\Lambda,}~\Omega_{m0}$. Solid line is for
$\Omega_{\Lambda}=0.80,~$dash-dash line is for $\Omega_{\Lambda}=0.70$ and the
dot-dot line is for $\Omega_{\Lambda}=0.60$. Note that $\Omega_{m0}%
=1-\Omega_{\Lambda}$. }%
\label{plot6}%
\end{figure}

In general it is not possible to write the potential, $V\left(  \phi\right)  ,
$ in closed-form. However, if we assume that $\phi_{1}=\phi_{0}=0$, then we
find that
\begin{equation}
\tau=-\frac{1}{3}W\left(  -2\frac{\Omega_{\Lambda}}{\Omega_{m_{0}}}e^{\frac
{9}{2}\phi}\right)  +\frac{3}{2}\phi\label{fttb.86}%
\end{equation}
which gives%
\begin{equation}
V\left(  \phi\right)  =4\Omega_{\Lambda}\left(  \frac{5}{2}-\left[  W\left(
-2\frac{\Omega_{\Lambda}}{\Omega_{m_{0}}}e^{\frac{9}{2}\phi}\right)  \right]
^{-1}-W\left(  -2\frac{\Omega_{\Lambda}}{\Omega_{m_{0}}}e^{\frac{9}{2}\phi
}\right)  \right)  , \label{fttb.87}%
\end{equation}
where $W$ is the Lambert Function. The qualitative behaviour of this potential
is given in Fig. \ref{plot6}.

\subsection{Solution in terms of the scalar field}

For completeness of our analysis we now consider the lapse function, $N\left(
t\right)  $, in which locally $\phi\left(  t\right)  =t$. In the following we
perform the change $t\rightarrow\phi$ and we express the solution in terms of
the scalar field.

Therefore, from the constrain equation (\ref{fttb.69}) we express the lapse
function $N\left(  \phi\right)  $ as%
\begin{equation}
N^{2}\left(  \phi\right)  =\frac{6h\left(  \phi\right)  \left(  h\left(
\phi\right)  -1\right)  }{V\left(  \phi\right)  } \label{fttb.88}%
\end{equation}
in which we substituted $a\left(  \phi\right)  =\exp\left(  \int h\left(
\phi\right)  dt\right)  $. Note that the $h\left(  \phi\right)  $ is not the
Hubble function, the latter is $H\left(  \phi\right)  =\frac{1}{\phi}h\left(
\phi\right)  $.

By using (\ref{fttb.88}) in (\ref{fttb.70}), or equivalently in (\ref{fttb.71}%
), we derive the first-order ordinary differential equation%
\begin{equation}
\frac{dh\left(  \phi\right)  }{d\phi}+\left(  1-h\left(  \phi\right)  \right)
h\left(  \phi\right)  \left(  \left(  6+2\xi\left(  \phi\right)  \right)
h\left(  \phi\right)  -\xi\left(  \phi\right)  \right)  =0, \label{fttb.89}%
\end{equation}
where now $V\left(  \phi\right)  =\exp\left(  \int\xi\left(  \phi\right)
d\phi\right)  .~$\ Equation (\ref{fttb.90}) is an Abel's Equation and can be
written as an algebraic equation. However, if we assume a specific function
$h\left(  \phi\right)  $, then we can calculate the corresponding potential by
solving the algebraic equation (\ref{fttb.89}) from which we can calculate
that
\begin{equation}
\xi\left(  \phi\right)  =\Xi\left(  h,\frac{dh}{d\phi}\right)  .
\label{fttb.90}%
\end{equation}

\section{Conclusions}

\label{conc}

This work was mainly focused on the existence and the stability for
cosmological exact relativistic solutions of special interest in a
higher-order modified teleparallel gravitational theory. The theory that we
considered belongs to the family of the $f-$theories in which the term which
modifies the Einstein-Hilbert Action depends upon the boundary term which
relates the invariants of the Levi-Civita and Weitzenb\"{o}ck connections.

The higher-order derivatives can be attributed to a noncanonical scalar field
with the use of a Lagrange Multiplier. This new scalar field is minimally
coupled with the Einstein tensors and the theory does not modify the
gravitational constant which we can say that the theory is defined in the
Einstein frame, in contrast with the plethora of higher-order modified
$f-$theories in which the scalar field equivalence in which the gravitational
constant becomes time-varying and the scalar field equivalence is defined in
the Jordan frame. In a similar way that one can work with O'Hanlon theory and
read the results in $f\left(  R\right)  $-gravity, we can work directly on the
scalar field description and extract results for the modified theory.

The main results which follow from our analysis are:

\begin{itemize}
\item The de Sitter Universe is an exact solution of the gravitational field
equation when for the noncanonical scalar field there exists at least a value
in which condition (\ref{fttb.30b}) is satisfied. The same condition can be in
terms of the $f$-theory and reads as the expression (\ref{fttb.30}) which is
the exact equivalent relation with the Barrow-Ottewill relation for $f\left(
R\right)  $-gravity. For different functions of the potential $V\left(
\phi\right)  $, there could be different points which describe the de Sitter
Universe. However, when $V\left(  \phi\right)  =V_{0}e^{-3\phi}$, condition
(\ref{fttb.30b}) is satisfied identically, while we found that for that
potential the de Sitter solution is an attractor. For other values of the
potential, $V\left(  \phi\right)  $, we found that the de Sitter solution can
be stable only when it describes an expanding Universe.

\item Furthermore in the case of a vacuum we derived the functional form of
the potential, $V\left(  \phi\right)  $, in order that the scalar field
behaves like an ideal gas. Specifically we found that $V\left(  \phi\right)
=V_{0}e^{-\lambda\phi},~\lambda\neq3$, where the equation of state parameter
is $\gamma=-2\left(  1-\frac{\lambda}{3}\right)  $, and scale factor $a\left(
t\right)  =a_{0}t^{\frac{2}{3\gamma}}$. That result is in agreement with that
of the analysis in the dimensionless variables \cite{anprd}. As far as
concerns the stability of that solution it was found that for $\lambda>0$ the
solution is stable. Therefore we observe that the theory can describe a dark
energy model with equation of state parameter lower than minus one.

\item We performed the same analysis in the presence of matter for which we
assumed two possible scenarios. Firstly, we assumed that the scalar field
mimics the matter source and there exists an exact solution if and only if the
scalar field potential is a power-law and the power is related with the
equation of state parameter for the matter source. The solution for $\gamma
\in\lbrack1,2)$ was found to be stable. For the second scenario we assumed
that the matter source dominates by providing the leading-order behaviour of a
singular power-law solution for the scale factor. That is a realistic scenario
as it can describe the matter-dominated era of the universe. In that scenario
we found an exact singular solution for the noncanonical scalar field and we
found that the stability of the solution depends upon the value for the power
of the power-law potential. There are two only possibilities, the singular
solution to be a stable or an saddle point and there is no any spiral (stable
or unstable) point. That result is different from that of the canonical scalar
field \cite{Liddle11}.
\end{itemize}

We continued our analysis by studying the degrees of freedom and the existence
of conservation laws for the field equations. We discussed the relation of
conservation laws which follows from the generators of Conformal
transformations for constraint Hamiltonian systems. Indeed, the field
equations form a constraint Hamiltonian system and the integrability was showed.

By using that latter discussion we were able to reduce the field equations to
one algebraic equation which is equivalent with the existence of algebraic
solutions, which is another way to define the integrability of differential
equations. We performed the reduction for simplicity in the case of vacuum for
two different frames and we described the solution in terms of the scale
factor or in terms of the noncanonical scalar field.

Furthermore we demonstrated our results by constructing some closed-form
solutions and we show how easily the Hubble function for the $\Lambda
$-cosmology can be reproduced from our theory in the case of a vacuum. Last,
but not least, we observed that dark matter components can be introduced by
the noncanonical scalar field into the evolution of the universe. That is
something which should be studied further.

There are various questions which should be answered. In a forthcoming work we
wish to extend our analysis by study the effects of this noncanonical scalar
field in the level of the perturbations and perform some cosmological constraints.

\begin{acknowledgments}
AP acknowledges the financial support of FONDECYT grant no. 3160121.
\end{acknowledgments}

\end{document}